\documentclass[twocolumn]{aastex631}
\usepackage{amsmath}
\allowdisplaybreaks[1]

\newcommand\brenna[1]{\textcolor{magenta}{[#1]}}

\shorttitle{Uncovering Hidden Massive Black Hole Companions with Tidal Disruption Events}
\shortauthors{Mockler et al.}
\graphicspath{{./}{figures/}}

\begin{document}

\title{Uncovering Hidden Massive Black Hole Companions with Tidal Disruption Events}


\correspondingauthor{Brenna Mockler}
\email{bmockler@carnegiescience.edu}

\author[0000-0001-6350-8168]{Brenna Mockler}
\affiliation{The Observatories of the Carnegie Institution for Science, Pasadena, CA 91101, USA}
\affiliation{Department of Physics and Astronomy, 
University of California, 
Los Angeles, CA 90095, USA}

\author[0000-0002-7854-1953]{Denyz Melchor}
\author[0000-0002-9802-9279]{Smadar Naoz}
\affiliation{Department of Physics and Astronomy,  
University of California, 
Los Angeles, CA 90095, USA}
\affiliation{Mani L. Bhaumik Institute for Theoretical Physics, Department of Physics and Astronomy, UCLA, Los Angeles, CA 90095, USA}

\author[0000-0003-2558-3102]{Enrico Ramirez-Ruiz}
\affiliation{Department of Astronomy and Astrophysics, University of California, Santa Cruz, CA 95064, USA}

\begin{abstract}

Dynamical perturbations from supermassive black hole (SMBH) binaries can increase the rates of tidal disruption events (TDEs). However, most previous work focuses on TDEs from the heavier black hole in the SMBH binary (SMBHB) system. In this work, we focus on the lighter black holes in SMBHB systems and show that they can experience a similarly dramatic increase in their TDE rate due to perturbations from a more massive companion. While the increase in TDEs around the more massive black hole is mostly due to chaotic orbital perturbations, we find that, around the smaller black hole, the eccentric Kozai-Lidov (EKL) mechanism is dominant and capable of producing a comparably large number of TDEs. In this scenario, the mass derived from the light curve and spectra of TDEs disrupted by the lighter SMBH companion is expected to be significantly smaller than the SMBH mass estimated from galaxy scaling relations, which are dominated by the more massive companion. This apparent inconsistency can help find SMBHB candidates that are not currently accreting as active galactic nuclei (AGN) and that are at separations too small to be resolved as two distinct sources. In the most extreme cases, these TDEs provide us with the exciting opportunity to study SMBHBs in galaxies where the primary SMBH is too massive to disrupt sun-like stars. 

\end{abstract}


\section{Introduction} \label{sec:intro}
Galaxy mergers are a natural consequence of hierarchical structure formation \citep[]{barnes_formation_2002, springel_formation_2005, robertson_merger-driven_2006, hopkins_cosmological_2008}.  Since nearly every galaxy hosts a supermassive black hole (SMBH), binary SMBHs (SMBHBs) should be common \citep[e.g.,][]{sillanpaa_oj_1988, rodriguez_compact_2006, komossa_recoiling_2008, bogdanovic_sdss_2009, boroson_candidate_2009, dotti_sdssj092712652943440_2009, Batcheldor:2010a, Deane:2014a, runnoe_large_2017,pesce_measuring_2018}. Observations of merging galaxies and of dual active galactic nuclei (AGN) give credence  to this idea \citep[e.g.,][]{Komossa:2004a, bianchi_chandra_2008, comerford_175_2009, comerford_origin_2018, Greene:2010a, liu_discovery_2010, smith_search_2010, stemo_catalog_2021, di_matteo_energy_2005, li_pairing_2020}, and it has even been suggested that our Galaxy might host a small companion to our own SMBH \citep[e.g.,][]{hansen_need_2003, maillard_nature_2004, gurkan_disruption_2005, gualandris_perturbations_2009, chen_is_2013, fragione_gravitational-wave_2020, gravity_collaboration_detection_2020, generozov_hills_2020, naoz_hidden_2020, zheng_influence_2020}

After two galaxies merge, the central supermassive black holes (SMBHs) will eventually sink to the bottom of the merged galaxy's potential, where they will form a binary system in the nucleus of the galaxy. 
While the two SMBHs coexist, they will induce orbital perturbations in each others' nuclear star clusters. These perturbations can send stars on high eccentricity, nearly radial orbits towards one of the black holes and lead to increased rates of tidal disruptions events \citep[e.g.][]{Ivanov:2005a, Chen:2011a, Wegg:2011a, Li:2015a, li_direct_2019, fragione_secular_2018}. 

Tidal disruption events (TDEs) occur when a star passes close enough to a black hole that the tidal force from the supermassive black hole (SMBH) overwhelms the self-gravity of the star \citep[e.g.][]{Rees:1988a,Evans:1989a}. The star is ripped apart and the stellar debris is accreted by the SMBH, producing a flare similar in brightness to supernovae that encodes properties of the star and SMBH in its light curve and spectra \citep[e.g.][]{Lodato:2009a, ramirez-ruiz_star_2009, guillochon_hydrodynamical_2013, Yang:2017,law-smith_tidal_2019, Mockler:2019a, mockler_evidence_2022}.
Theoretical tidal disruption event rates have most often been calculated using the dynamics of two-body relaxation \citep[e.g.,][]{magorrian_rates_1999, wang_revised_2004,MacLeod:2012a, stone_rates_2016}. In this picture, a star will experience random kicks to its angular momentum through interactions with other stars, and if it is very unlucky, one of these kicks will send it radially toward the SMBH, where it will be tidally disrupted. However, in a SMBH binary, 3-body interactions can produce TDEs much faster than two-body relaxation processes \citep[e.g.][]{Ivanov:2005a, Chen:2011a, Wegg:2011a}. 


In general, SMBH binary systems are very difficult to uncover observationally. If one or both of the black holes is accreting as an AGN, it is sometimes possible to measure the shift in velocity in the emission lines due to the binary orbit \citep{gaskell_quasars_1983, boroson_candidate_2009, decarli_peculiar_2010, tsalmantza_systematic_2011, eracleous_large_2012, runnoe_large_2015, wang_searching_2017}. There are also predictions for how the overall shape of the broad emission lines might differ from a solitary AGN \citep[]{decarli_peculiar_2010, nguyen_emission_2016}. Unfortunately, measuring the velocity shift in emission lines is only possible over a limited range of SMBHB separations. The SMBHs must be close enough such that the velocity shifts in the broad line regions (BLR) are measureable, but not so close that the BLRs are truncated \citep[see overview in][]{de_rosa_quest_2019}. 

If we are lucky enough to catch both SMBHs actively accreting, this is likely the surest sign of a SMBH binary (SMBHB) system, however, resolving these systems requires the SMBHs to still be at large separations (before the system has formed a bound binary) and are also mostly limited to binaries with mass ratios near $q=1$ \citep[e.g.,][]{rodriguez_compact_2006, comerford_origin_2018, dey_unique_2019, goulding_discovery_2019, bhattacharya_automated_2020, foord_second_2020, kim_dual_2020, reines_new_2020, monageng_radio_2021, severgnini_possible_2021, shen_hidden_2021}. 
Additionally, even though the merger process can funnel gas to the center of galaxies and trigger AGN activity, most SMBHBs in the local universe do not currently host AGN \citep[][]{casey-clyde_quasar-based_2022}. 


When stars are disrupted by the smaller SMBH in the binary, the black hole mass measured from the TDE light curve or spectra will correspond to the smaller SMBH \citep[e.g.][]{Mockler:2019a, ramsden_bulge_2022, wen_continuum-fitting_2020, ryu_measuring_2020}. 
Given that the potential of the galactic nucleus will be dominated by the larger SMBH, typical scaling relations are likely to produce estimates close to the mass of the larger black hole in the binary \citep[]{izquierdo-villalba_unveiling_2022}. Therefore, if TDE light curve measurements suggest a much smaller black hole mass than scaling relations such as $M-\sigma$ or the bulge-mass relation, this is an indication that the galaxy might host a SMBH binary\footnote{Note that if the SMBHB is on a very close orbit such that the orbital timescale is comparable to the fallback timescale of tidal debris, other observables such as truncations in the accretion rate can also become important \citep[e.g.][]{Mardling:2001a, Ricarte:2016a}}. Similarly, if a TDE is observed in a galaxy that appears to be too massive to host an observable flare (i.e. a system with an expected black hole mass $\gtrsim 10^8 {\rm M}_\odot$), this also suggests the existence of a second, smaller SMBH in the galactic nucleus \citep[e.g.][]{coughlin_tidal_2018}. 

In this paper, we study how hierarchical 3-body interactions between the SMBH binary and the stars orbiting the smaller black hole can excite the eccentricities of the stars and increase the rates of tidal disruption events around the smaller black hole. In particular, we focus on the effect of the secular orbital perturbations known as the eccentric Kozai-Lidov (EKL) mechanism \citep{kozai_secular_1962, lidov_evolution_1962, naoz_eccentric_2016}.  
It was shown that SMBH binaries and the EKL mechanism can 
lead to increased rates of TDEs. 
However, most previous work on this subject has either focused on the stars surrounding the larger SMBH \citep[e.g.][]{Ivanov:2005a, Chen:2011a, Wegg:2011a, fragione_calibrating_2021}, on IMBH systems \citep[]{fragione_secular_2018}, or on extreme mass ratio inspiral (EMRI) transients instead of TDEs \citep[][]{naoz_combined_2022, mazzolari_extreme_2022}.
Past work has also shown that other dynamical mechanisms suppress EKL oscillations for stars around the larger black hole, and the majority of TDEs that are produced around the larger SMBH come from chaotic orbital crossings \citep[e.g.,][]{Chen:2011a}. This is generally not the case for the stars around the smaller SMBH, largely because the EKL perturbations due to the more massive SMBH  produce stronger excitations in the stars' orbital eccentricities. Here we expand on work by \citet{Li:2015a} and explore the effect of the EKL mechanism on the rates of TDEs around the smaller SMBH in SMBH binaries.

We propose that every 
SMBH binary system will undergo a burst (short, enhanced period) of TDEs from the smaller black hole once the binary separation is of order the sphere of influence (so long as one of the SMBHs is below the Hills mass for the star, $\approx 10^8 {\rm M}_\odot$). At these separations, the binary will not be resolved spatially, and the TDE will appear to be at the center of the galaxy. TDEs from the smaller black hole provide a unique avenue to discover SMBHBs at much closer separations than the kiloparsec scales at which dual AGN are observed. This is especially significant when the more massive SMBH is unable to disrupt sun-like stars outside its event horizon \citep[as also discussed in][]{fragione_secular_2018} .


\section{Methodology}\label{sec:method}

\begin{figure}[ht!]
\begin{center}
\includegraphics[width = \columnwidth]{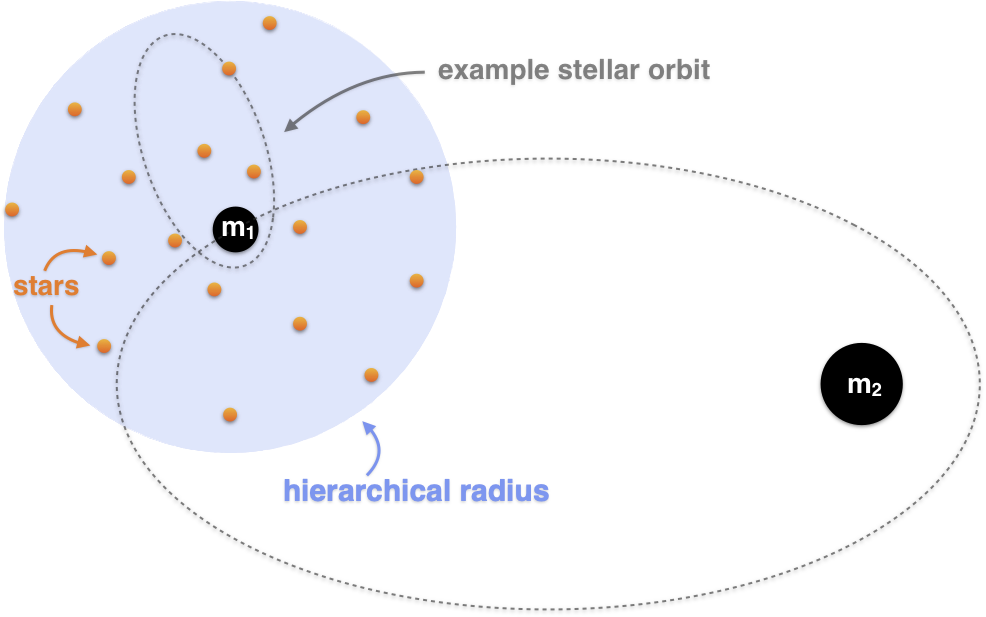} 
\end{center}
\vspace{-0.5cm}
\caption{ Cartoon of simulation setup (not to scale). We study stars orbiting the smaller of the two SMBHs, and focus on the region within the hierarchical radius (defined in Equation~\ref{eq:a_hierarchical}), as orbits in this region are less likely to be chaotically scattered by the second black hole. 
\label{fig:cartoon}
}
\end{figure}

\subsection{Physical processes}

We consider a SMBH binary $m_1 < m_2$, separated by a semi-major axis $a_{\rm bin}$, and with an eccentricity $e_{\rm bin}$. Throughout the paper, the subscripts `1', `2', and `*', denote the inner black hole, the outer black hole, and the star, respectively. The subscript `bin' refers to the black hole binary. Surrounding each SMBH is a nuclear star cluster, and for the purposes of this paper we will focus on the stars surrounding $m_1$. Because $m_1 < m_2$, we can think of the nuclear star cluster of $m_1$ as embedded in the larger cluster around $m_2$. We expect the smaller SMBH $m_1$ to retain stars within its Roche lobe \citep[e.g.][]{oleary_recoiled_2012, rantala_formation_2018, greene_search_2021}. 
Each star in the cluster around $m_1$ can be modeled as part of a hierarchical three-body system made up of an inner binary pair -- $m_*$ and $m_1$ --  and an outer binary pair -- $m_1$ and $m_2$ (see illustration in Figure~\ref{fig:cartoon}). We assume the stars orbiting $m_1$ follow approximately Keplerian orbits, but over time their orbits can be secularly perturbed by $m_2$ through the eccentric Kozai-Lidov mechanism \citep[e.g.][]{naoz_eccentric_2016}. 

\begin{figure}[ht!]
\centering 
\includegraphics[width = \columnwidth]{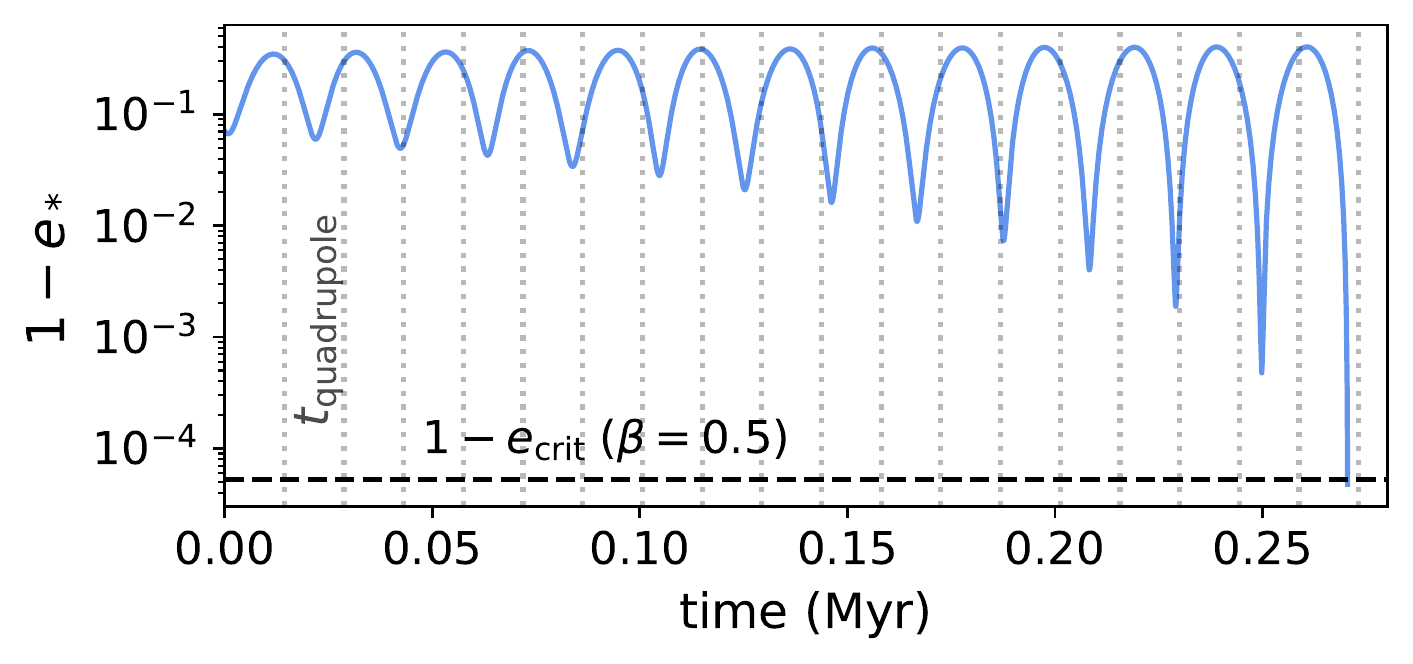} 
\caption{Example run showing EKL cycles as a function of time. The dashed horizontal line denotes our critical radius for partial disruption ($\beta = 0.5$), and the dotted vertical lines denote $t_{\rm quad}$ given the initial orbital conditions. As you can see, $t_{\rm quad}$ is of order the period of one eccentricity oscillation, as expected. This simulation is part of the set of runs with $m_1 = 10^6{\rm M_\odot}, \; m_2 = 10^7{\rm M_\odot},\; \alpha = 2$. 
}
\label{fig:EKL_cycle_vs_time}
\end{figure}

We restrict the initial parameter space to stable configurations, adopting a hierarchical limit that ensures the orbit of the star stays well within the orbit of the binary.
\begin{equation}\label{eq:epsilon}
    \epsilon = \frac{a_*}{a_{\rm bin}} \frac{e_{\rm bin}}{1 - e_{\rm bin}^2} < 0.1 \ ,
\end{equation} 
where $\epsilon$ is the pre-factor in front of the octupole expansion term of the Hamiltonian \citep[e.g.,][]{Naoz:2013a}\footnote{This limiting condition was found by \citet [][]{Naoz:2013b} to give results that are similar to the more complicated \citet[][]{Mardling:2001a} stability criterion.}
This defines a `hierarchical radius' limit:
\begin{equation}\label{eq:a_hierarchical}
    a_{\rm hier} = 0.1 \times a_{\rm bin} \frac{1-e_{\rm bin}^2}{e_{\rm bin}} \ .
\end{equation}

The EKL mechanism describes the coherent perturbations from $m_2$ on the stars orbiting $m_1$. These perturbations exchange angular momentum between the outer SMBH binary ($m_1$ and $m_2$) and the inner binaries made up of stars orbiting $m_1$. This produces excitations in the eccentricity and inclination of the stars orbiting $m_1$ over a wide range of the parameter space \cite[e.g.,][]{naoz_hot_2011, katz_long-term_2011,lithwick_eccentric_2011,naoz_secular_2013,li_eccentricity_2014, will_orbital_2017}. Importantly, the eccentricity excitations can lead to the production of a large number of tidal disruption events \citep[e.g.,][]{Li:2015a, fragione_secular_2018}. To model these systems and determine the rate of tidal disruption events, we employ the secular approximation, where the inner and outer orbits are treated as phase-averaged wires that torque each other.  
We solve the secular, hierarchical three-body equations following \citet{Naoz:2013a}, including up to the octupole order terms ($\mathcal{O}(a_{*}/a_{\rm bin})^3$). The timescale associated with the quadrupole level of approximation is \citep[e.g.,][]{antognini_timescales_2015}: 
\begin{equation}
 t_{\rm quad}  =  \frac{16}{15}\frac{a_{\rm bin}^3 (1 - e_{\rm bin}^2)^{3/2} \sqrt{(m_1 + m_*)}}{\sqrt{G} a_*^{3/2} m_2}  \ , \label{eq:tkl}
\end{equation}
where $G$ is the gravitational constant. This timescale is associated with the period of oscillations of the eccentricity and inclination (see Figure~\ref{fig:EKL_cycle_vs_time}). 

\begin{figure*}[ht!]
\begin{center}
\includegraphics[width = \textwidth]{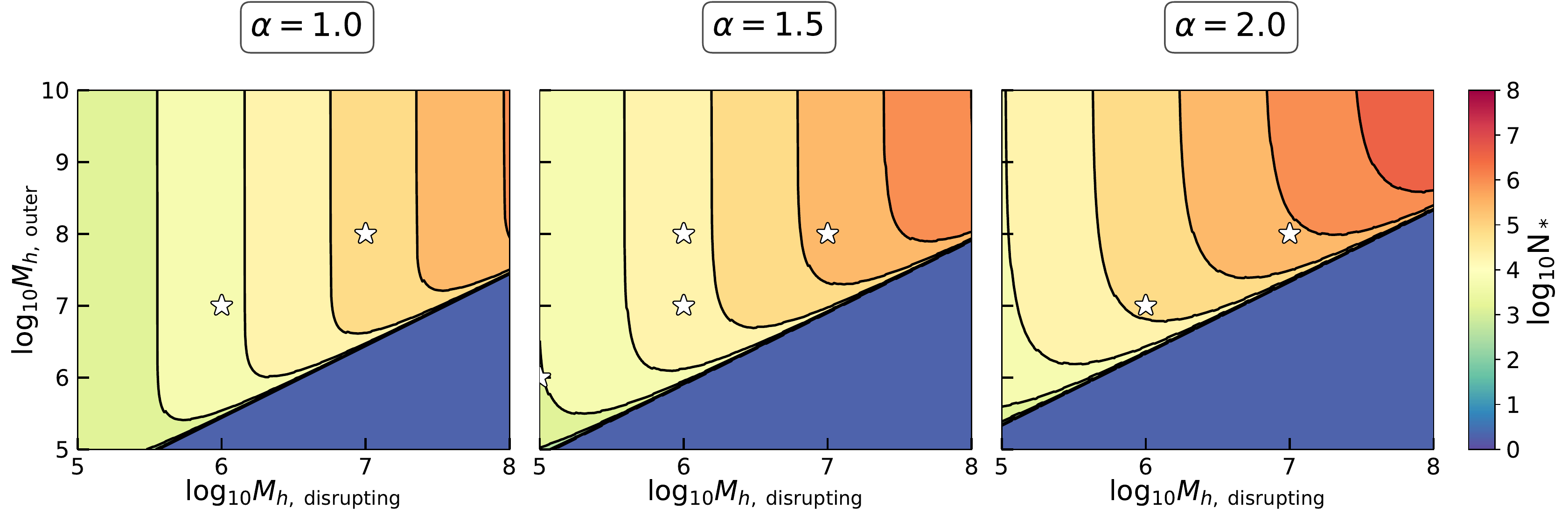} 
\end{center}
\caption{Analytical exploration of the parameter space showing the number of stars susceptible to the EKL mechanism. The white stars represent the parameter combinations chosen for numerical simulation runs. This was calculated by determining the range of semi-major axes for each parameter combination where the EKL timescale is the shortest dynamical timescale. All of our systems have $e_{\rm bin} = 0.5$, $a_{\rm bin} = 0.5 r_{h, \; \rm disrupting}$. While we compared against a variety of dynamical timescales (including those for two-body relaxation, vector resonant relaxation, general relativistic precession, and binary hardening through gravitational wave emission, see Appendix~\ref{sec:timescales}), 
the two competing timescales were $t_{\rm GR, \; 1}$ and $t_{\rm NT}$. These are the timescales for general relativistic (GR) precession due to the disrupting black hole and Newtonian precession due to the stellar potential, respectively.  
}
\label{fig:paramspace_Ntde}
\end{figure*}


\citet{Li:2015a} considered the relevant dynamical timescales at the center of galaxies and compared them to the EKL timescale. They found that the dynamical processes competing with the EKL mechanism for the largest effect on the stellar orbital parameters are
general relativistic precession (due to the potential of the disrupting black hole), and Newtonian precession (due to the potential of the stellar cusp). The timescale for general relativistic precession is,
\begin{equation}
t_{\rm GR1}  =  \frac{2 \pi a_*^{5/2} c^2 (1 - e_*^2)}{3 G^{3/2} (m_1 + m_*)^{3/2}} \ , \label{eq:tGR1}
\end{equation}
where $c$ is the speed of light. 
The Newtonian (NT) precession timescale is given by:
\begin{equation}
t_{\rm NT}  = 2 \pi \left( \frac{\sqrt{G m_1/a_*^3}}{\pi m_1 e_*} \int_0^\pi {\rm d} \psi \: M_\ast(r) \: {\rm cos} \psi  \right)^{-1} \ ,  \label{eq:tNT}\\
\end{equation} 
where $\psi$ is the true anomaly of the inner orbit, $r(\psi) = a_*(1 - e_*^2)/(1 + e_* {\rm cos} \psi)$ from Kepler’s equation. The function $M_\ast (r)$ is the integrated stellar mass within $r(\psi)$ -- we will discuss the density profiles used to calculate this in Section~\ref{sec:MC}. If precession due to GR or the stellar potential is strong enough, it can prevent EKL perturbations from building up coherently and exciting the eccentricity and inclination of the star \citep[e.g.,][]{Li:2015a,Chen:2011a}. 


\begin{table*}[ht]
\centering
\caption{Simulation parameters and results. See Section~\ref{sec:stability} for a description of how upper and lower limits were calculated for the rightmost three columns.}
    \renewcommand{\thefootnote}{\arabic{footnote}}
    \footnotesize
    \setlength\tabcolsep{3pt}
    \renewcommand{\arraystretch}{1.2}
\begin{tabular}{ccccccccccc}
\hline
     run \# & $m_1$ & $m_3$ & $\alpha$ & $a_1$ max & $N_{*}$ & TDE fraction & $N_{TDE}$ & $N_{plunge}$ \\
     & $\rm (log_{10} {\rm M}_\odot)$&$\rm (log_{10} {\rm M}_\odot)$ & & (au) & & & & & &\\
     
\hline


0 & 5.0 & 6.0 &  1.50 &  6002 &    5134 &        0.11 -  0.44 &       576 -      2277 &           0 -           5 \\
1 & 6.0 & 7.0 &  1.00 & 18980 &   14081 &        0.07 -  0.45 &       986 -      6379 &           0 -          70 \\
2 & 6.0 & 7.0 &  1.50 & 18980 &   51307 &        0.09 -  0.41 &      4618 -     20984 &          51 -         257 \\
3 & 6.0 & 7.0 &  2.00 & 18980 &  182705 &        0.09 -  0.33 &     15713 -     59562 &           0 -         731 \\
4 & 6.0 & 8.0 &  1.50 & 18980 &   51366 &        0.01 -  0.43 &       462 -     22242 &           0 -         308 \\
5 & 7.0 & 8.0 &  1.00 & 60021 &  140784 &        0.05 -  0.38 &      7180 -     54061 &         141 -        3801 \\
6 & 7.0 & 8.0 &  1.50 & 60021 &  512509 &        0.05 -  0.32 &     23575 -    165028 &        1025 -       10250 \\
7 & 7.0 & 8.0 &  2.00 & 60021 & 1860848 &        0.05 -  0.27 &    100486 -    504290 &           0 -       20469 \\
\hline
\end{tabular}
\label{table:simparams}
\end{table*}

Motivated by a similar analysis in \citet{Li:2015a}, in Figure~\ref{fig:paramspace_Ntde}, we look at how changing  different parameters of the system changes the number of stars likely to experience large eccentricity oscillations from the EKL mechanism. For each point in Figure~\ref{fig:paramspace_Ntde}, we calculate the number of stars within our hierarchical limit 
(Equation~\ref{eq:a_hierarchical}) whose shortest dynamical timescale is $t_{\rm quad}$, and plot density contours displaying this information. 
The dynamical timescales we compare against are described in the figure caption (and included in Appendix~\ref{sec:timescales}). Like \citet{Li:2015a}, we find that $t_{\rm GR1}$ and $t_{\rm NT}$ are the two shortest timescales apart from $t_{\rm quad}$, implying that general relativistic precession due to the disrupting black hole and Newtonian precession due to the stellar cusp are the two dynamical processes most likely to interfere with EKL oscillations. This parameter space exploration also further motivates our focus on the smaller black hole in the binary. The number of stars susceptible to EKL oscillations drops off sharply as the disrupting black hole grows to a comparable size to the outer black hole \citep[previous work focusing on the larger black hole found that the majority of disruptions were caused by chaotic orbital crossings, not the EKL mechanism, e.g.,][]{Wegg:2011a,Chen:2011a}.  

Orbital precession due to GR and the stellar cusp 
can be important at small radii, however the strength of EKL perturbations 
increase with a star's radius from $m_1$ as stars move closer to the perturbing black hole ($m_2$). 
Therefore it is more difficult for the EKL mechanism to excite the eccentricities of stars at smaller semi-major axes, but at larger radii, where there are many more stars, the EKL mechanism often dominates and can produce many TDEs. For this work we chose to isolate the effects of the EKL mechanism and used the parameter space exploration described by Figure~\ref{fig:paramspace_Ntde} to choose systems to model where 
the EKL timescale $t_{\rm KL}$ is the shortest dynamical timescale for the majority of stars surrounding the smaller black hole.

We include GR precession in our modeling, and while it is possible that it suppresses the number of disruptions at small radii close to $m_1$ (see Figure~\ref{fig:fTDE}), we find that the EKL mechanism acts much faster than GR precession for the majority of the stars around $m_1$ in the systems we are modeling, as expected \citep[see for similar results][]{naoz_formation_2014,Li:2015a,naoz_combined_2022}. 
Unlike \citet{Li:2015a}, we do not include the NT precession from the stellar cusp in our models, and instead chose our setups such that the timescale for NT precession is longer than the timescale for EKL excitations for the majority of the stars. Thus, our results are consistent with \citet{Li:2015a} (see Section~\ref{sec:results}). Finally, we note that we do not include any star-star collisions or compact object collisions. We expect that these processes will generally become important on longer timescales \citep[e.g.,][]{rose_formation_2022,rose_stellar_2023}.

Another possible dynamical mechanism that can affect EKL perturbations is two-body relaxation. 
\citet[][]{naoz_combined_2022} showed that even though the canonical timescale for two-body relaxation to significantly change stellar orbits is many orders of magnitude larger than the Kozai-Lidov timescale in the relevant parameter space (consistent with our calculations for Figure~\ref{fig:paramspace_Ntde}), individual (incoherent) kicks from two-body scattering can affect the build up of Kozai-Lidov perturbations. In this paper we focus on the effects of the EKL mechanism alone, however, in a companion paper we explore the effect of two-body relaxation on the EKL mechanism  \citep[see][]{melchor_tidal_2023}.

\subsection{Monte Carlo dynamical simulations}\label{sec:MC}

Using Figure~\ref{fig:paramspace_Ntde} as our guide, we chose a variety of parameter combinations for our simulations that we expected to produce observable tidal disruption events. The parameters we varied were: the mass of the disrupting and perturbing black holes ($m_1$ and $m_2$), their mass ratio ($q = m_2/m_1$), 
and the power law constant of the density profile for the stellar cusp ($\alpha$). 
The mass of $m_1$ was set to a range of reasonable values for tidal disruption events: $10^5, 10^6,$ and $10^7$ solar masses. Black holes above $\approx 10^8 {\rm M}_\odot$ will swallow most stars whole, and disrupt them within their event horizons. Black holes below $10^5 {\rm M}_\odot$ can and do disrupt stars \citep{Law-Smith:2017a}, however these disruptions may be more difficult to observe \citep[e.g. due to slower circularization and lower Eddington limits,][]{Guillochon:2015b}, and almost all confirmed TDEs have occurred around black holes above $10^5 {\rm M}_\odot$ (e.g., \citealt{ramsden_bulge_2022, hammerstein_final_2022}, with the exception of \citealt{angus_fast-rising_2022}). Therefore, we focus on this mass range for the purposes of this paper (see \citealt{fragione_secular_2018} for simulations of similar systems focusing instead on IMBH-SMBH binaries). We use mass ratios $q = m_2/m_1$ of 10 and 100, and this determines the mass of $m_2$.

The stellar cusp is modeled using the density distribution described in \citet{OLeary:2009a}: 
\begin{equation}\label{eq:density}
    \rho_\ast (r) = \frac{3 - \alpha}{2\pi} \frac{m_1}{r^3} \left( \frac{G M_0 (m_1/M_0)^{1 - 2/k}}{\sigma_0^2 r}\right)^{-3 + \alpha} \ ,
\end{equation}

and our sphere of influence is defined where the mass enclosed is equal to $2\times m_1$,

\begin{equation}\label{eq:rh}
r_h = \frac{G M_0 (m_1/M_0)^{1-2/k}}{\sigma_0^2} \ .
\end{equation}

The density profile is written as a power law of semi-major axis, with the integrated stellar mass normalization fixed by the $M-\sigma$ relation. The constants $M_0 = 1.3 \times 10^8 {\rm M}_\odot$, $\sigma_0 = 200 ~\rm km ~s^{-1}$, and $k=4$ are the best-fit values for the $M-\sigma$ relation in \citet{tremaine_slope_2002} ($M_h/{\rm M}_\odot = M_0 (\sigma/\sigma_0)^k$). We vary the power law exponent `$\alpha$' between 1-2 to explore how the steepness of the stellar cusp affects TDE rates. While our own galactic center appears to have a flatter cusp \citep[$\alpha \sim 1$, e.g., ][]{schodel_distribution_2018}{}{}, TDE host galaxies tend to be very centrally concentrated and to have stellar light profiles indicative of steep stellar cusps down to the size scales that are observable \citep[e.g.,][]{law-smith_tidal_2017,french_structure_2020,Dodd:2021}. 

\begin{figure}[ht!]
\centering 
\includegraphics[width = \columnwidth]{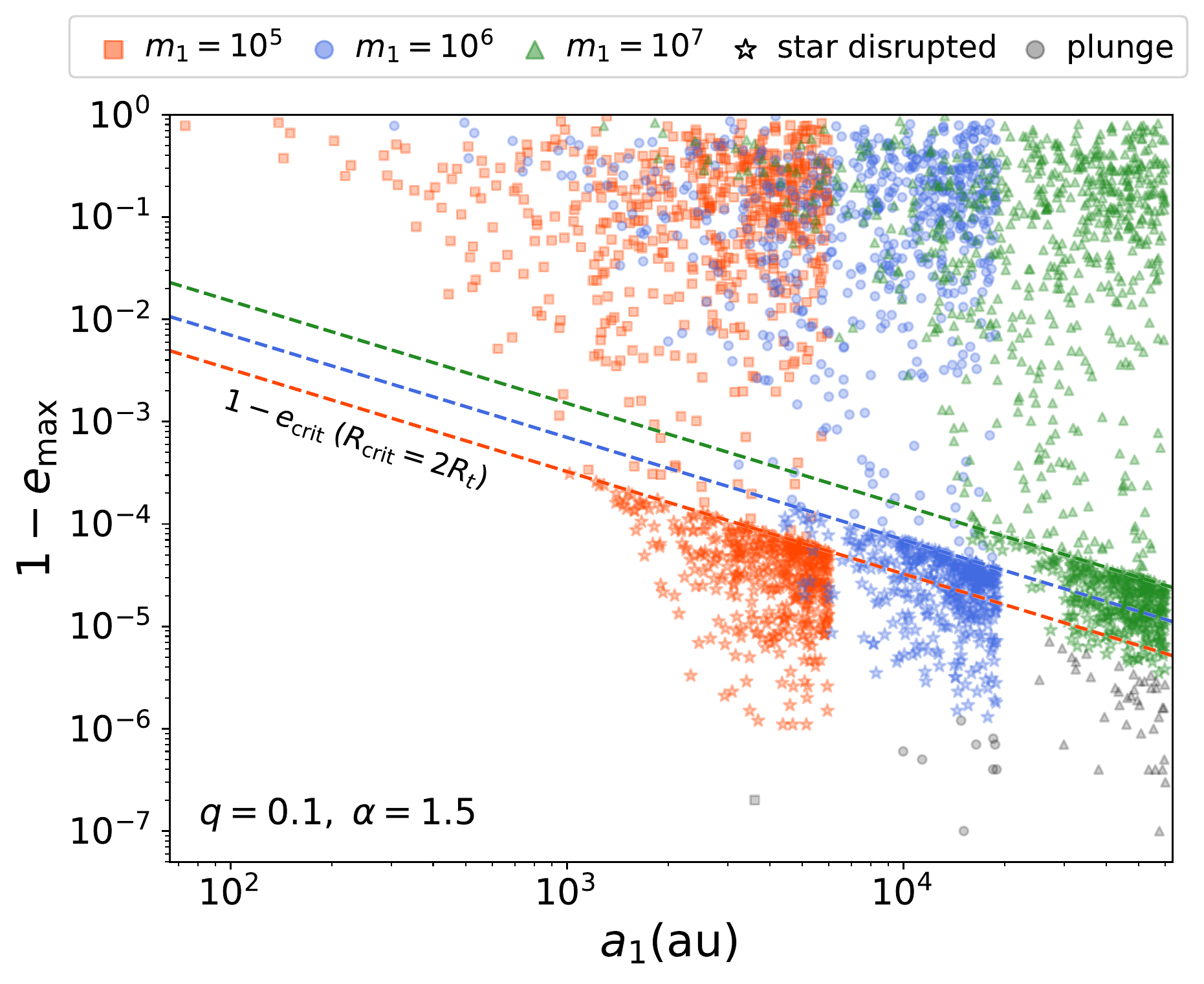} 
\vspace{-0.5cm}
\caption{The final distributions of the stars' semi-major axes ($a_1$) and maximum eccentricities ($e_{\rm max}$, plotted as $1 - e_{\rm max}$) for 3 runs with varying inner black hole mass ($m_1 = 10^5, 10^6, 10^7 {\rm M}_\odot$). The critical radius for disruption used here is $R_p \leq 2R_T$ (allowing for both partial and full disruptions), and denoted by dashed lines in the plot. Disruptions are plotted with a star-shaped marker, and plunges (disruptions that occur with $R_T < R_S$) are shaded in gray. 
The maximum semi-major axis included for stars in a given run is defined by the hierarchical condition (Equation~\ref{eq:epsilon}).}
\label{fig:a1_emax_scatter}
\vspace{-0.1cm}
\end{figure}

We set the eccentricity of the SMBH binary to an intermediate eccentricity of $e_{\rm bin} = 0.5$ and its semi-major axis to half of the sphere of influence of the disrupting black hole ($a_{\rm bin} = 0.5 r_{h, \;1}$). This puts the SMBHs at separations of $\sim 1$~pc ($a_{\rm bin} =0.2 - 1.9$~pc for $m_1 = 10^5 - 10^7 {\rm M}_\odot$) --  a regime where the EKL mechanism is highly effective, however the SMBH binary's hardening timescale is still long enough that we do not have to worry about the binary's orbit changing significantly over the course of our simulations. At these radii, the SMBH binary's hardening timescale is mainly dependent on the effectiveness of loss-cone diffusion, which is likely $\gtrsim 100$ million years \citep[e.g.][]{kelley_massive_2017}, much longer than the EKL timescales for the majority of the stars in our simulation setups. 

For each set of parameter combinations, we ran a 1000 numerical simulations of the 3-body interactions with different initial conditions for the star. The priors for the star's initial conditions were determined as follows: The semi-major axis of the star was drawn randomly from the density distribution, with a maximum radius determined by the hierarchical condition of the system (see Equation~\ref{eq:epsilon}). The star's eccentricity was drawn from an isothermal distribution, and the mutual inclination was drawn from an isotropic distribution (uniform in cos($i$)). The argument of periapsis and the longitude of the ascending nodes were drawn from a uniform distribution between $0-2\pi$. See Table~\ref{table:simparams} for an overview of the simulation parameters. 

The simulations were run until either the star was disrupted or the system was evolved for $10^9$ years (by which point the TDE rate from the EKL mechanism had long since dropped off, see Figure~\ref{fig:TDErate}). Our criteria for a tidal disruption to occur was that the star passed within twice the tidal radius ($ R_p = 2 R_T$) of the black hole
\begin{equation}
    R_T = R_* \Big( \frac{M_h}{M_*}\Big)^{1/3} \ .
\end{equation}
This meant that our minimum impact parameter for disruption was $\beta = R_T/R_p = 0.5$. 
Setting our TDE radius condition to $2R_T$ instead of $R_T$ ensured that we captured partial disruptions as well as full disruptions\footnote{Below $\beta \approx 0.5$ very little mass will return to the black hole and it is unlikely the TDE will be observable \citep[e.g.,][]{Guillochon:2013b}}.

Each set of a 1000 simulations provides a random Monte Carlo sampling of the stars around the smaller black hole in each binary system, given the priors described above. We plot the final eccentricity distributions of a subset of our runs in Figure~\ref{fig:a1_emax_scatter}, where the runs that end in disruption are clearly visible as a buildup of points at high eccentricities (low values of $1-e$). 

Our simulations give us the fraction of stars that result in TDEs for each parameter combination, as well as the time of disruption for each random sample. We also find that a number of stars plunge directly into the black hole, with pericenter radii less than the Schwarszchild radii of the black holes (shaded in gray in Figure~\ref{fig:a1_emax_scatter}).
We note that if the SMBH is spinning, the star may zoom around the ergosphere instead of plunging directly into the black hole \citep[e.g.,][]{ glampedakis_zoom_2002,healy_zoom-whirl_2009,schnittman_distribution_2015}.  This can continue for a few tens to thousands of times the pericenter passage \citep[see figure 3 in][]{naoz_dark_2019}. 

Finally, we can scale our results for the fraction of stars disrupted by the expected total number of stars around $m_1$ within the hierarchical radius (using Equation~\ref{eq:density}) to compare to real galaxies. The simulation parameters and the fraction and number of stars that become TDEs are quoted in Table~\ref{table:simparams} for each of our runs. We can also calculate a TDE rate as a function of time by binning our simulation results as a function of the time of disruption. We plot our scaled, time-dependent rates for a subset of the runs in Figure~\ref{fig:TDErate} (using Gaussian processes to smooth the distributions).

\subsection{Limits and stability}\label{sec:stability}

For each set of simulations, we calculate an upper and lower limit for the TDE fraction. The lower limit assumes that any star whose orbit evolves such that its apocenter moves outside the Hill radius of the smaller black hole is scattered away from the smaller black hole and is not disrupted. We define the Hill radius at the black hole binary's pericenter to ensure our lower limit is sufficiently conservative,
\begin{equation}
    r_{\rm Hill} = a_{\rm bin} (1 - e_{\rm bin})\left(\frac{m_1}{3m_2}\right)^{1/3} \ , \label{rhill}
\end{equation}
The upper limit assumes that all stars (within our hierarchical radius limit, see Equation~\ref{eq:a_hierarchical}) on disrupting orbits become TDEs, even those whose apocenters move outside the Hill radius before disruption. While the Hill radius is a good approximation of the stability boundary on long timescales \citep[e.g.,][]{grishin_generalized_2017, bhaskar_mildly_2021, tory_empirical_2022}, stars at larger radii have shorter EKL timescales, and can easily be disrupted before their orbits become unstable. Stars whose apocenters are just outside the Hill radius are not instantaneously scattered -- the impulse felt on the orbit from the second black hole must build up over time before the star's orbit is significantly affected \citep{zhang_stability_2023}.

It is also possible for stars to remain ``Lagrange-stable'' even when they are no longer ``Liapunov-stable'' \citep[e.g.,][]{hayashi_lagrange_2022}, meaning that stars can perform chaotic figure eight orbits around both black holes but remain bound to the system. In this situation, a star can become unbound from the first SMBH only to return later and get disrupted. 

Finally, three-body scattering experiments have shown that even when stars are far outside the Hill radius, significant fractions of stars can be disrupted through chaotic orbital crossings with the second SMBH \citep[although this has mostly been studied for stars orbiting the larger black hole, e.g.][]{Chen:2011a}. Given that we are focused here on studying TDEs disrupted through the EKL mechanism, we only consider disruptions within the hierarchical radius, which is within a factor of two of the Hill's radius in our simulations with $q=10$ (and within a factor of four for simulations with $q=100$), even for eccentricities approaching 1.

Thus, we conclude that using the Hill radius as a lower limit produces estimates on the TDE rate that are too conservative, and the true rate is somewhere in between the limits used in this paper. We note that a similar approach was taken in \citet{naoz_combined_2022} and \citet{melchor_tidal_2023}. See the former study's appendix for an exploration of the particles' trajectories.  

\section{Results}\label{sec:results}

In this section we present the results of our dynamical simulations, focusing on the rates and fraction of stars that become tidal disruption events.

\subsection{Rates}\label{rates}
\begin{figure}[ht!]
\begin{center}
\includegraphics[scale = 0.45]{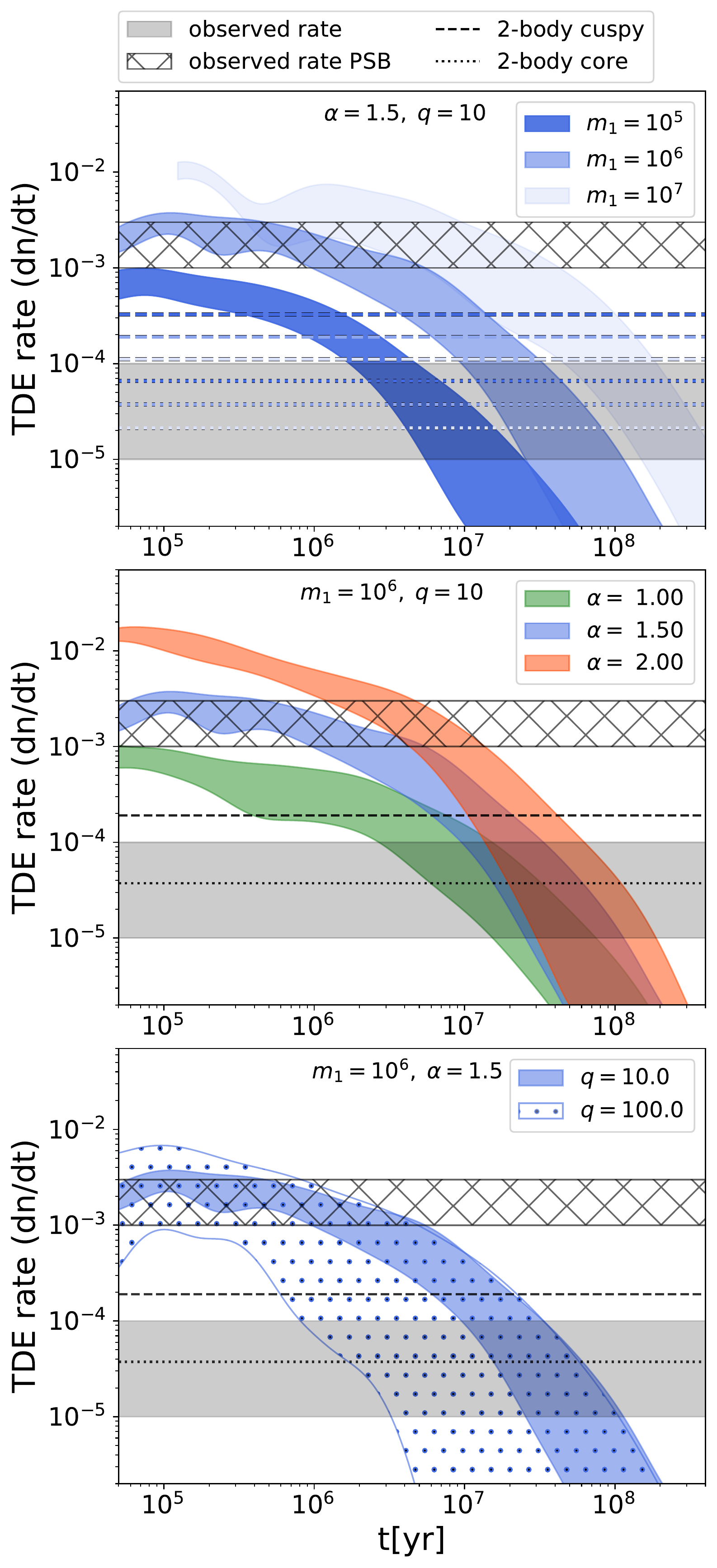} 
\end{center}
\vspace{-0.5cm}
\caption{TDE rates calculated for the EKL mechanism compared to rates calculated for 2-body relaxation and to observed rates.  The 2-body relaxation rates plotted for `cuspy' galaxies (dashed lines) correspond to galaxies with an inner density profile ($\rho \propto r^{-\alpha}$) $\alpha > 0.5$, while the rates plotted for `core' galaxies (dotted lines) correspond to $\alpha < 0.3$ \citep[][note that density profiles are estimated observationally from radii $> r_h$, and therefore are not a direct comparison to the density profiles used in simulations in this paper for radii interior to $r_h$.]{Stone:2016a}.
The observed rate for all TDEs (shaded in gray) is from \citet{van-Velzen:2020}, and the observed rate in post-starburst galaxies (hatched region) is from \citet{French:2020a}.
\label{fig:TDErate}
}
\end{figure}

As you can see in Figure~\ref{fig:TDErate}, we obtain the expected `burst' of TDEs in all of our simulations.
The bursts peak around $\sim 10^5 - 10^6$ years but remain higher than the global observed rate ($\lesssim 10^{-4}$ tdes/galaxy/year) for $\sim 10^6 - 10^8$ years. This is comparable to or longer than the burst timescales predicted in \citet[]{Chen:2011a} for the more massive black hole companion (e.g. comparing our timescales for $10^6 M_\odot$ secondaries to \citet[]{Chen:2011a} for $10^7 M_\odot$ primaries).
The burst is limited by the eccentric Kozai-Lidov timescale in that it will not begin until the stars' eccentricities have grown large enough for disruptions to occur. However, the burst can extend much longer than the eccentric Kozai-Lidov timescale (which is generally of order $10^4 - 10^6$ years for these systems), as stars often go through many EKL cycles before they are disrupted.

We also note that we do not include the evolution of the black hole binaries in these simulations. 
If we modeled them as they evolved, we would expect the EKL `loss-cone' to move from larger radii to smaller radii, as stars at larger radii are disrupted or scattered and the perturbations grow stronger at small radii. Therefore, we would likely expect a more extended rate if we included this binary evolution. 
This is what \citet{Chen:2011a} found in their n-body simulations which included binary hardening due to stellar scattering \citep[and][for similar simulations focused on EMRIs]{mazzolari_extreme_2022}{}. Initially, while the disruption timescale was comparable to the hardening timescale, the peak of the burst was extended, but then when the hardening timescale grew longer than the tidal disruption timescale and the loss cone could no longer be replenished, the rate dropped off (although the authors note that loss-cone
diffusion processes and gravitational wave emission were not included, and would likely result in further hardening at later times and a longer-lived plateau in the rate). 

\subsection{TDE fraction}\label{sec:fractions}

The fraction of stars that become tidal disruption events depends on the  number of stars within the hierarchical limit and the strength of the EKL mechanism i.e., the maximum value of the  eccentricity excitations. In the test particle limit, the eccentricity can be excited to extreme values \citep[e.g.,][]{li_eccentricity_2014, naoz_formation_2014}, naturally driving the stars to low angular momentum orbits. 
We find that for the majority of our simulations, the upper limit on the percentage of stars that are disrupted is $\approx 30-45\%$ (within the maximum radius defined by our hierarchical condition, Equation~\ref{eq:a_hierarchical}). This limit does vary slightly with the parameters chosen. 

\begin{figure}[ht!]
\begin{center}
\includegraphics[scale = 0.52]{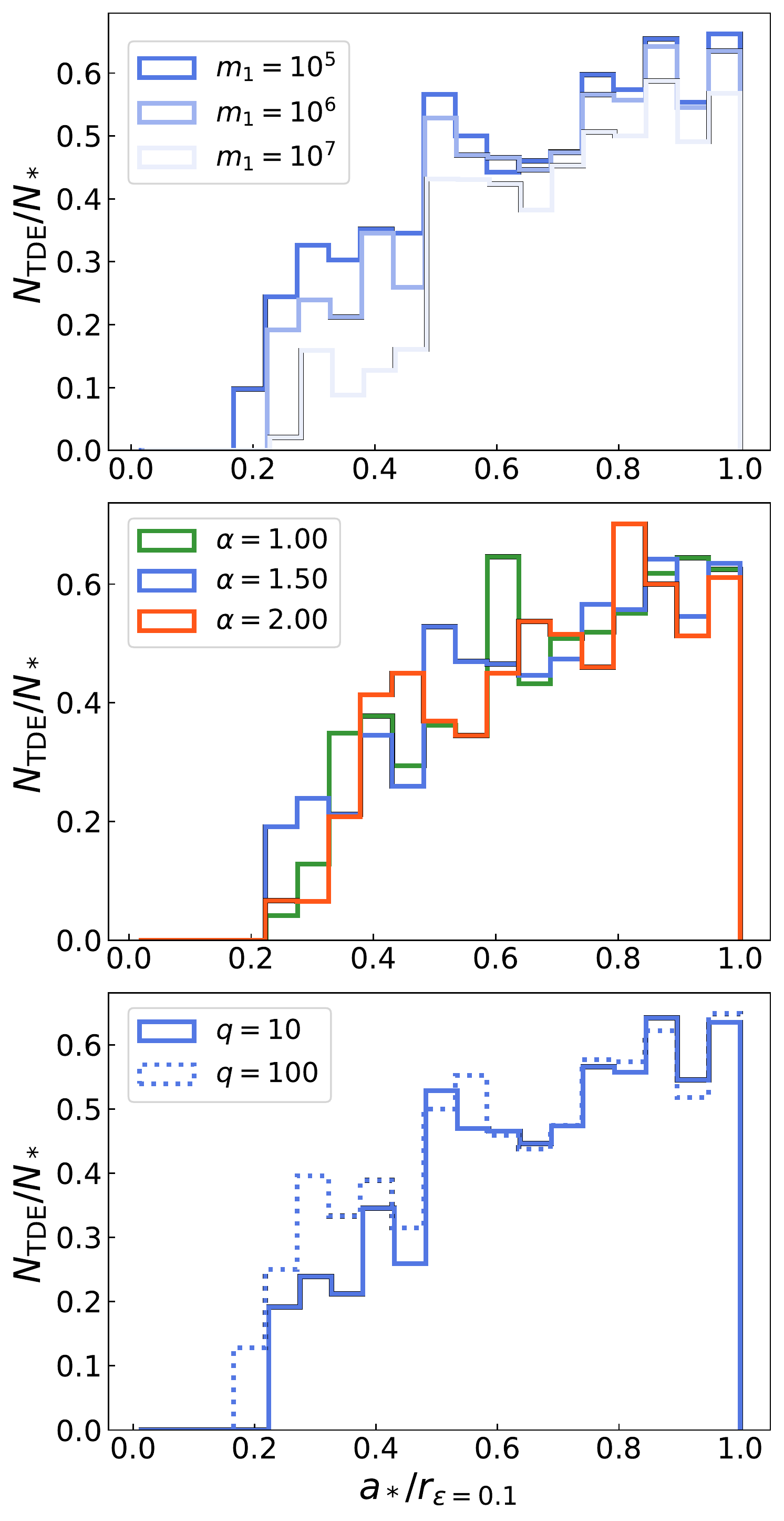} 
\end{center}
\vspace{-0.25cm}
\caption{ The fraction of stars that are disrupted as TDEs as a function of semi-major axis (scaled by the hierarchical radius). Each panel varies one simulation parameter. {\bf Top:} $\alpha = 1.5$, $q=10$. {\bf Middle:} $m_1 = 10^6 {\rm M}_\odot$, $q=10$. {\bf Bottom:} $m_1= 10^6 {\rm M}_\odot$, $\alpha = 1.5$.
\label{fig:fTDE}}
\end{figure}

As depicted in Figure~\ref{fig:fTDE}, the fraction of stars disrupted decreases as $m_1$ increases, particularly at small radii. This is mostly because GR precession is faster than the EKL precession  
close to the black hole. On the other hand, the fraction of stars disrupted at small radii increases when $m_2$ increases because the strength of the EKL oscillations increases with the mass of the perturber. The number of disruptions at large radii (close to the hierarchical radius) does not vary significantly because we are effectively running out of stars at those radii. Near the hierarchical radius, most of the stars in the inclination range susceptible to EKL perturbations are getting excited and disrupted, so increasing the strength of the perturbations is not noticeably increasing the number of disruptions. As there are much fewer stars at small radii ($M_{*, \rm contained} \propto r^{3-\alpha}$), increasing the mass ratio to 100 only increases the upper limit on the total fraction of disrupted stars in our simulations by $\approx 2\%$ (see Table~\ref{table:simparams}).

\begin{figure*}[ht!]
\begin{center}
\includegraphics[width = \textwidth]{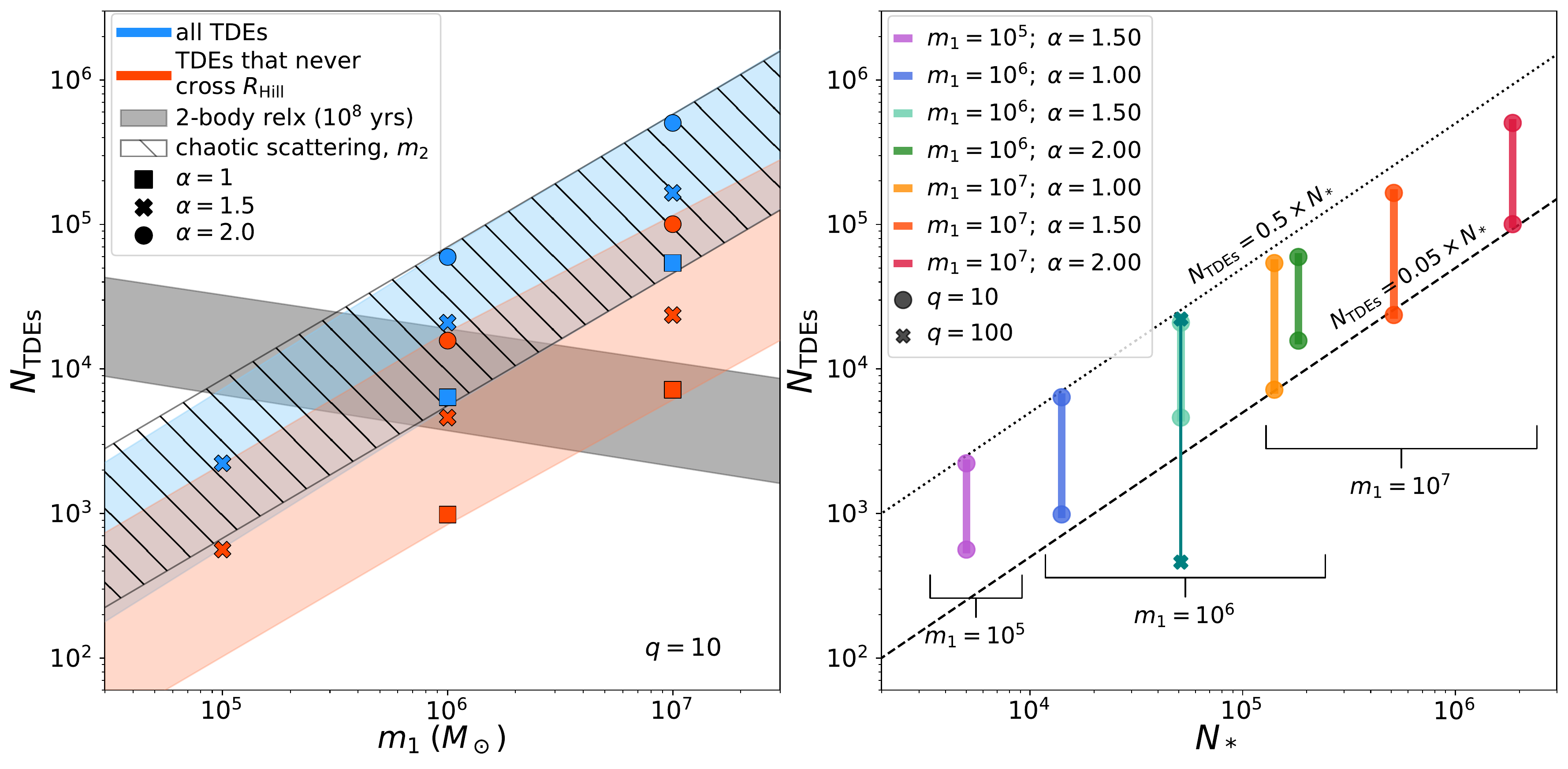}
\vspace{-0.1cm}
\caption{{\bf LHS:} The number of TDEs as a function of $m_1$, with different density profiles represented as different marker shapes (q=10 in all cases). The points in orange are the lower limits, excluding all stars that move outside $r_{\rm Hill}$, while the points in blue are the upper limits and include these stars. The blue and orange shaded regions are plotted by determining the slopes between the fraction of TDEs for the $\alpha = 1 \; \& \; 2$ runs at the black hole masses used in the simulations and then extrapolating out to higher and lower masses. 
The black hatched region is the number of TDEs disrupted by $m_2$ in the same system from chaotic scatterings (the upper limit is from \citealt{Liu:2013a} and the lower limit is scaled down to match \citealt{Wegg:2011a}). The gray shaded region is the number of TDEs predicted from 2-body relaxation over $10^8$ years for a single SMBH system with mass $m_1$ \citep[from][]{stone_rates_2016}.
{\bf RHS:} The number of TDEs as a function of the number of stars within the hierarchical radius. Each $m_1$ and $\alpha$ combination is plotted as a different color, and the marker shape is determined by the mass ratio $q$. The upper and lower limits are determined in the same way as for the LHS.
} 
\label{fig:Ntde}
\end{center}
\end{figure*}

As expected, the density profile power law ($\alpha$) affects the maximum fraction of TDEs, because changing $\alpha$ changes the relative fraction of stars near the hierarchical radius, where most of the disruptions originate.  
The number of stars disrupted is proportional to $1/t_{\rm KL}$, which is, in turn, proportional to $a_*^{3/2}$. This can be used to estimate the upper limit on the total fraction of stars disrupted. The number of stars disrupted as a function of radius is dependent on both the eccentric Kozai-Lidov timescale and the density profile, and therefore goes as
\begin{equation}
    n_{\rm disrupt}(r) \propto 1/t_{\rm KL} \times n_*(r) \propto  r^{3/2}r^{2-\alpha}.
\end{equation}
Then, the total fraction of stars disrupted goes as the integral of $n_{\rm disrupt}(r)$ divided by the total number of stars
\begin{equation}
    \frac{\int{n_{\rm disrupt}(r) dr}}{ \int{n_*(r) dr}} \propto \frac{\int{r^{3/2}  r^{2-\alpha} dr}}{\int{r^{2-\alpha} dr}} = \frac{r_{\rm max}^{3/2} ( 3 - \alpha)}{(9/2 - \alpha)},
\end{equation}
and so the upper limit on the total fraction of stars disrupted (integrated over the radius range) is proportional to:
\begin{equation}
    \rm{TDE \; fraction \; (upper \; limit)} \propto \frac{3-\alpha}{9/2 - \alpha}.
\end{equation}
While changing the mass of both black holes and the steepness of the stellar cusp affects the fraction of stars disrupted, it does not change the maximum TDE fraction by more than $\sim 0.1$ (see Table~\ref{table:simparams}). However, changing these parameters does dramatically change the number of stars within the hierarchical radius (our maximum radius, see Equation~\ref{eq:epsilon}), and therefore the number of stars that can become TDEs. Because of this, we find that the strongest determinant of the maximum number of TDEs in the system is simply the number of stars within the hierarchical radius ($N_*(r_{\rm max}) \propto r_{\rm max}^{9/2 - \alpha}$)\footnote{We note that if we did not choose simulation parameters such that $t_{KL} < t_{NT}$ and $t_{KL} < t_{GR1}$ for the majority of stars, the steepness of the stellar cusp and the mass of the black holes might have a more noticeable affect on the fraction of stars that become TDEs.}. This is shown clearly in Figure~\ref{fig:Ntde}. 

On the other hand, the lower limit on the number of TDEs is strongly dependent on the mass ratio $q$. The lower limit on the number of TDEs is proportional to the number of stars that are not susceptible to scattering by the perturber -- stars that remain within the Hill radius ($r_{\rm Hill} \propto q^{-1/3}$). The number of stars within the Hill radius follows the proportionality $N_*(r_{\rm Hill}) \propto r_{\rm Hill}^{9/2 - \alpha}$.

Combined with the fact that $r_{\rm Hill} \propto q^{-1/3}$, the lower limit on the TDE fraction goes as\footnote{We note that the TDE fractions are scaled by the number of stars within $r_{\rm hier}$, and $r_{\rm hier}$ is not dependent on the mass ratio `q'. If we instead calculated the lower limit on the TDE fraction by scaling by the number of stars within $r_{\rm Hill}$, the fraction would go as $\propto r_{\rm Hill}^{3/2} \propto q^{-1/2}$ instead.},
\begin{equation}
    \rm{TDE \; fraction \; (lower \; limit)} \propto q^{\frac{\alpha}{3} - \frac{3}{2}} \ . 
\end{equation}
It drops from $\approx 10\%$ for $q=10$ to $\approx 1\%$ for $q=100$ ($m_1 = 10^6$, $\alpha = 1.5$). 
We note that we find similar qualitative results to that of \citet{Li:2015a} (see following section), however, here we explore a wider part of the parameter space. 

As we discussed in the previous section, our lower limit on the stability radius is likely too conservative. Our lower limit is approximately equal to the strongest limits found in \citet[][]{grishin_generalized_2017} and \citet[][]{tory_empirical_2022} ($a_* \lesssim 0.5r_H$). However, some of the stars outside this radius are disrupted on very short timescales, likely before their orbits would become unstable. Recently, \citet{zhang_stability_2023} determined an analytical stability criterion for 3-body systems that they found to be consistent with n-body simulations. 
Adopting the \citet{zhang_stability_2023} stability timescale definition means that stars from larger radii can undergo EKL eccentricity excitations before becoming unstable. This would increase the lower limit on the number of TDEs by a factor of $\approx 2$ across all of our simulations.
Therefore, it is likely that many of these stars will be disrupted before their orbits become unstable. It is also likely that many of the stars on unstable orbits will also become TDEs, even though analytical approaches are unable to model their evolution (as discussed in Section~\ref{sec:stability}).



\subsection{Comparison with previous work}

We compared our results with \citet{Li:2015a}, as they used a similar setup to our own. 
In their simulation, $m_1 = 10^7 {\rm M}_\odot$, $m_2 = 10^8 {\rm M}_\odot$, $\alpha = 1.75$, $e_2 = 0.5$ and $a_2 = 0.5$pc, and they find 5.7\% of stars within the Hill's radius are disrupted (using the more lenient criteria that a star must only pass within $3 R_T$ to be disrupted). 
This is comparable to the results for our simulations numbered 5-7, which were run with similar parameters and found that $\approx 5\%$ of stars within the Hill's radius are disrupted. To compare directly, we also ran one additional test simulation with the same parameters (and disruption criteria) as \citet{Li:2015a}, and found that 4.2\% of stars within the Hill's radius were disrupted. 
We note that \citet{Li:2015a} include the effect of orbital precession due to the stellar cusp in their simulations but that our simulations appear consistent nonetheless, justifying our exclusion of this effect for this set of simulations (see Section~\ref{sec:method} for further explanation). 

We also compared our results to the 3-body scattering simulations of the eccentric Kozai-Lidov mechanism around the larger SMBH in \citet{Chen:2011a} and \citet{Wegg:2011a}. Both of these works showed that the EKL oscillations induced by a smaller SMBH on stars surrounding a larger companion are suppressed for large parts of the parameter space. 
In Figure~\ref{fig:TDErate}, we show that we get a comparable number of TDEs from the smaller black hole in our $q=10$ simulations as predicted for the larger black hole by scaling relations calculated from the results in \citet{Chen:2011a} and \citet{Wegg:2011a}.  

\section{Discussion}\label{sec:discussion}


We find that the EKL mechanism in SMBH binaries produces a burst of tidal disruption events at rates much higher than those expected from two-body relaxation (see Figure~\ref{fig:TDErate}). This phenomenon has been previously explored for tidal disruption events, but the majority of past work has focused on the disruptions of stars orbiting the more massive black hole in the binary system \citep[e.g.][]{Ivanov:2005a, Chen:2009a, Chen:2011a, Wegg:2011a}. This meant that the majority of disruptions did not actually come from eccentric Kozai-Lidov oscillations, but rather from chaotic orbital crossings between the stars and the perturbing black hole \citep{Chen:2011a, Wegg:2011a}. As we are interested in determining when eccentric Kozai-Lidov oscillations in particular are important, we followed \citet{Li:2015a}'s lead and explored a parameter space centered on the smaller black hole.

\subsection{Connections to simulated and observed TDE rates}

In Figure~\ref{fig:TDErate} we compare the rate calculated from our simulations with the rates estimated for two-body relaxation based on properties of observed galaxies \citep[dashed lines, from][]{Stone:2016a}. We find that the rate from the EKL mechanism is higher than the rate from two-body relaxation for timescales between millions to tens of millions of years depending on the parameters of the system. This means that during a burst the EKL rate is also much higher than the overall observed rate (as two-body relaxation calculations generally over-predict the observed rate).  However, the scenario described in this paper is only applicable to a subset of galaxies, and only over a relatively short time period.

\subsubsection{Caveats}
To begin with, only a fraction of galaxies will host close SMBH binaries at any given time. Given the uncertainties on SMBHB hardening timescales \citep[e.g.][]{kelley_massive_2017}, it is not straightforward to estimate the current population of SMBH binaries from the rate of galaxy mergers, however, an overview is still useful for understanding possible scenarios. 

The current day merger fraction (as defined by the fraction of galaxies with close companions) is $\lesssim 3\%$ for galaxies $\lesssim 10^{11} M_\odot$ \citep[this is the major merger fraction, the minor merger fraction is likely similar,][]{bundy_greater_2009, lopez-sanjuan_dominant_2012}. 
However, it is possible that the current-day SMBHB population reflects a higher merger fraction from  earlier times, depending on the hardening timescale of the binary (\citealt{bluck_surprisingly_2009, bluck_structures_2012} find a major merger fraction of up to $\approx 25 - 40\%$ at $z=3$, which is $\approx 10^{10}$ year lookback time). 
If the SMBHB population is reflective of these higher merger rates, upwards of 25\% of galaxies might host two SMBHs at a given time \citep[]{rodriguez-gomez_merger_2015}, but their relevance to this work would still depend on their hardening timescales.

In addition to requiring that the host galaxies and their SMBHs be at a specific evolutionary stage, 
the actual burst of TDEs from the EKL mechanism is short-lived. 
If the stars are not replenished rapidly through star formation or dynamical mechanisms, the rate will drop off relatively quickly and the elevated rate will only be observable for tens of millions of years for anything but the most massive systems (see Figure~\ref{fig:TDErate}). Therefore, it is possible that even in galaxies with close SMBH binaries, we will miss the `burst' of TDEs predicted here.

Finally, we note that although we focus on the tidal disruptions around the smaller of the black holes in the SMBH binary, the global rate will be limited by the number of larger mass SMBHs available to form binaries because the black hole mass function decreases with increasing mass \citep[e.g.][]{shankar_selection_2016,gallo_exploring_2019}.


\subsubsection{Comparison with two-body relaxation}
Given these restrictions, it is reasonable that we do not observe dramatically higher TDE rates due to the EKL mechanism. However, we still expect this mechanism to contribute meaningfully to TDE rates, particularly for higher mass systems with lower rates from two-body relaxation. 
In Figure~\ref{fig:Ntde}, the number of TDEs from two-body relaxation from a single galaxy is comparable to the number from an EKL burst over 100 million years for a disrupting black hole mass of $10^6$ (perturbing $M_h = 10^7 M_\odot$). At higher black hole masses, the EKL mechanism dominates even over these relatively long timescales. While this comparison is still during the bursting phase, \citet{stone_delay_2018} estimated the average TDE rates per galaxy for SMBHBs (assuming that the larger black hole dominated the rates) and compared it to two-body relaxation rates. They found that the average rate from SMBHBs approaches the rate from two-body relaxation for black hole masses $\gtrsim 10^{7.5} M_\odot$. When we compare the number of disruptions from our simulations with the number of disruptions expected around the larger black hole (from chaotic orbital interactions), we find that the numbers are similar for $q=10$ \citep[see hatched region in Figure~\ref{fig:Ntde},][]{Chen:2011a, Wegg:2011a}. Assuming the smaller black hole contributes a similar number of disruptions would mean that the average rate from SMBHBs would approach the rate from two-body relaxation for primary SMBH masses around $10^{7} M_\odot$. In addition to increasing the total number of disruptions from each SMBHB pair, it would also allow for the inclusion of higher mass galaxy mergers where the larger black hole is past the Hill mass turnover point  ($\approx 10^8 M\odot$).

\subsubsection{Comparison with post-starburst rates}
We note that the rate we calculate during the EKL burst is much more comparable to the observed rates calculated for post-starburst galaxies, which are often considered candidates for recent mergers \citep[as mergers can trigger starbursts]{Mihos:1994, Bekki:2005, Hopkins:2009a}. The reason why the TDE rate in post-starburst galaxies is $\sim 20-200 \times$ higher than the galaxy averaged rate remains an open question in the field \citep[e.g.,][]{Arcavi:2014a, Law-Smith:2017a, French:2020a,Dodd:2021}. Previous work has suggested that SMBH binaries might be able to increase the rates of TDEs in these galaxies if the starburst is due to a recent merger \citep[e.g.][]{Arcavi:2014a}. However, as \citet{Stone:2018a} point out, if the starburst is triggered by a merger, and the increased rates are from the resulting SMBH binary, this requires the smaller black hole to sink to the center of the galaxy on timescales short enough that the post-starburst features are still observable. If the starburst is triggered when the merger starts, this may be difficult, as the timescales for most SMBH binaries to harden to distances of $\approx 1$pc are thought to be of order $10^9$ years \citep[][]{kelley_massive_2017}. Additionally, the observed TDEs from post-starburst galaxies appear to have SMBH masses measured from light curves that are consistent with the SMBH masses from galaxy scaling relations \citep{Mockler:2019a, ramsden_bulge_2022}. Therefore, even if the galaxies host SMBH binaries, it is unlikely that the smaller black hole is responsible for the disruptions and therefore that the EKL mechanism is producing these TDEs. 


\subsection{Finding hidden SMBH binaries}
Despite their rarity, bursts of TDEs produced by the EKL mechanism could help us find hidden SMBH binaries. Tidal features from minor mergers dissipate quickly \citep{conselice_asymmetry_2000}, and the gravitational potential of a galactic nucleus hosting a SMBH binary with a mass ratio $\gtrsim 10$ will be dominated by the mass of the larger black hole, making it difficult to find SMBH binary candidates. However, as shown here, disruptions produced by the EKL mechanism will produce a significant number of TDEs around the smaller black hole in a SMBH binary, comparable in number to TDEs around the larger black hole for mass ratios of $q=10$ (see Figure~\ref{fig:Ntde}). Tidal disruption flares contain information about the mass of the disrupting black hole, therefore TDEs from the smaller black hole can expose its presence. For example, the light curve timescale of a TDE is dependent on the mass of the black hole \citep[e.g. {$t \propto M_h^{1/2}$},][]{Mockler:2019a}, therefore if a very short flare occurred around what seemed to be a much larger black hole, that would be an indication of a potential SMBH binary. If the mass of the larger black hole is above $\gtrsim 10^8 {\rm M}_\odot$, it would not be able to produce an observable flare from the disruptions of most stars, and therefore seeing any TDE at all from very massive galaxies would be a very strong indication of a hidden companion SMBH \citep[e.g.][]{coughlin_tidal_2018, fragione_secular_2018}.

\section{Summary of Key Findings}
Our key results are summarized as follows:

\begin{itemize}
    \item The EKL mechanism produces a burst of TDEs around the smaller black hole in a SMBHB at a rate that is orders of magnitude higher than the rate from two-body relaxation for timescales of millions to tens of millions of years (see Figure~\ref{fig:TDErate}). Unlike EKL oscillations around the larger SMBH, the EKL oscillations around the smaller SMBH are not significantly suppressed by precession due to general relativity or the stellar cusp. This is mostly because of the increased strength of the perturber. 
    \item The number of disruptions is primarily dependent on the number of stars within the hierarchical (or Hill) radius. This is because the fraction of stars disrupted does not change dramatically with most parameters (so long as we choose regions of parameter space such that the EKL mechanism is the dominant dynamical mechanism, see Figure~\ref{fig:Ntde}, Figure~\ref{fig:paramspace_Ntde}, Table~\ref{table:simparams}). The timescale is most strongly dependent on the black hole mass and mass ratio (see Figure~\ref{fig:TDErate}).
    \item The EKL mechanism generates a larger fraction of disruptions around the smaller black hole ($\sim 3\% - 45\% $ of stars within the hierarchical radius are disrupted) than chaotic orbital interactions are able to produce around the larger black hole \citep[e.g.][]{Chen:2011a, Wegg:2011a}. 
    This can lead to a comparable number of disruptions around both black holes (see Figure~\ref{fig:Ntde}).
    \item Disruptions around the smaller black hole can provide evidence of the existence of the SMBH binary as their light curve timescales will likely be shorter than what is expected from the larger black hole, which dominates the potential \citep[][]{Mockler:2019a}.
\end{itemize}

We have shown that tidal disruptions from the secondary black hole in a SMBHB contribute meaningfully to the rate from the binary system. We emphasize that this provides a novel way to find SMBHB candidates at separations that are too small to resolve distinct sources. 

\begin{acknowledgments}
We thank the participants and organizers of the fall 2023 Como meeting on the dynamics of TDEs and EMRIs for fruitful discussions and helpful clarification. 
B.M. is grateful for support from the U.C. Chancellor's Postdoctoral fellowship and the Mani L. Bhaumik Institute for Theoretical Physics as well as from Swift (80NSSC21K1409). D.M. is grateful for support from the NSF graduate fellowship. S.N. acknowledges the partial support from NASA ATP 80NSSC20K0505 and from NSF-AST 2206428 grant as well as thanks Howard and Astrid Preston for their generous support. E.R.R. thanks the Heising-Simons Foundation, NSF (AST-1615881 and AST-2206243), Swift (80NSSC21K1409, 80NSSC19K1391) and Chandra (22-0142) for support. The calculations for this work were carried out in part on the UCSC lux supercomputer (supported by NSF MRI grant AST-1828315).
\end{acknowledgments}

%

\vspace{5mm}

\software{astropy \citep{Astropy-Collaboration:2013a} 
          }



\appendix

\section{Dynamical timescales\label{sec:timescales}}

We present an overview of the relevant dynamical timescales for our SMBH binary systems. These were used in the making of Figure~\ref{fig:paramspace_Ntde}.

\begin{align}
& t_{\rm quad}  =  \frac{16}{15}\frac{a_{\rm bin}^3 (1 - e_{\rm bin}^2)^{3/2} \sqrt{(m_1 + m_*)}}{\sqrt{G} a_*^{3/2} m_2} & \\
& t_{\rm GR, \: 1}  =  \frac{2 \pi a_*^{5/2} c^2 (1 - e_*^2)}{3 G^{3/2} (m_1 + m_*)^{3/2}}  & \\
& t_{\rm GR, \: 2}  =  \frac{2 \pi a_{\rm bin}^{5/2} c^2 (1 - e_{\rm bin}^2)}{3 G^{3/2} (m_* + m_1 + m_2)^{3/2}} &   \\ 
& t_{\rm GR, \: int} =  \frac{16}{9}\frac{a_{\rm bin}^{3} c^2 (1 - e_{\rm bin}^2)^{3/2}}{ \sqrt{a_*}e_* \sqrt{1 - e_*^2} G^{3/2} m_1^{1/2} m_2} & \\
& t_{\rm NT}  = 2 \pi \left( \frac{\sqrt{G m_1/a_*^3}}{\pi m_1 e_*} \int_0^\pi {\rm d} \psi \: M_\ast(r) \: {\rm cos} \psi  \right)^{-1} & \label{eq:tNT2} \\
& t_{\rm RR, \: v} =  \frac{2 \pi f_{\rm vrr}}{\Omega} m_1 \frac{1}{\sqrt{M_\ast(r) m_*}} & \label{eq:tRRv}\\
& t_{\rm rel} =  0.34 \frac{\sigma_\ast^3}{G^2 \rho m_* {\rm ln} \Lambda} &  \label{eq:trel} \\
& t_{\rm LT} = \frac{a_*^3 c^3 (1 - e_*^2)^{3/2}}{2 G^2 m_1^2 s} & \label{eq:tLT} \\
& t_{\rm GW} = \frac{a_{\rm bin}^4}{4} \frac{5}{64} \frac{c^5}{G^3 m_1 m_2 (m_1 + m_2)} &
\end{align}

The equations above represent timescales for the following physical processes: 
$t_{\rm quad}$ is the quadrupole eccentric kozai-lidov timescale \citep{naoz_secular_2013, antognini_timescales_2015};
$t_{\rm GR, \: 1}$ \& $t_{\rm GR, \: 2}$ are the first-order post-Newtonian GR precession timescales for the inner and outer black holes; and $t_{\rm GR, \: int}$ is the timescale of the first order post-Newtonian GR interaction between the inner and outer orbits. As defined in  \citet[][]{kocsis_resonant_2011}, $t_{\rm NT}$ is the Newtonian precession timescale caused by the potential of the stellar cusp, $t_{\rm RR, \: v}$ is the vector resonant relaxation timescale, and $t_{\rm LT}$ is the Lense-Thirring precession timescale. Finally, $t_{\rm GW}$ is the binary SMBH orbital decay timescale due to gravitational wave emission. In equation~\ref{eq:tRRv}, $\Omega$ is the orbital frequency of the star and $f_{vrr}$ is estimated to be 1.2 by \citet{kocsis_numerical_2015}.  In equation~\ref{eq:tNT2}, $\psi$ is the true anomaly of the inner orbit, $r(\psi) = a_1(1 - e_1^2)/(1 + e_1 cos \psi)$ from Kepler’s equation. The function $M_\ast (r)$ is the integrated stellar mass within $r(\psi)$. 
In equation~\ref{eq:trel}, $\rm ln \Lambda$ is the Coulomb logarithm, and $\sigma_\ast$ is the velocity dispersion of the bulge. 
In equation~\ref{eq:tLT}, the constant $(G^2 m_0^2 s)/c$ is the spin angular momentum of the inner SMBH \citep{Peters:1964a, kocsis_resonant_2011, naoz_resonant_2013}.





\bibliography{zotero_library, library}{}
\bibliographystyle{aasjournal}



\end{document}